\def\fracm#1#2{\hbox{\large{${\frac{{#1}}{{#2}}}$}}}

\def\lin{\vrule width0.5pt height5pt depth1pt}
\def\dpx{{{ =\hskip-3.75pt{\lin}}\hskip3.75pt }}

\documentstyle[12pt]{article}

\skewchar\fivmi='177
\skewchar\sixmi='177
\skewchar\sevmi='177
\skewchar\egtmi='177
\skewchar\ninmi='177
\skewchar\tenmi='177
\skewchar\elvmi='177
\skewchar\twlmi='177
\skewchar\frtnmi='177
\skewchar\svtnmi='177
\skewchar\twtymi='177
\def\@magscale#1{ scaled \magstep #1}
\skewchar\fivsy='60
\skewchar\sixsy='60
\skewchar\sevsy='60
\skewchar\egtsy='60
\skewchar\ninsy='60
\skewchar\tensy='60
\skewchar\elvsy='60
\skewchar\twlsy='60
\skewchar\frtnsy='60
\skewchar\svtnsy='60
\skewchar\twtysy='60

\catcode`@=11
\def\un#1{\relax\ifmmode\@@underline#1\else
        $\@@underline{\hbox{#1}}$\relax\fi}
\catcode`@=12

\def\d{\delta}
\def\e{\epsilon}
\def\f{\phi}

\def\l{\lambda}

\def\o{\omega}

\def\s{\sigma}

\def\z{\zeta}

\def\L{\Lambda}

\def\S{\Sigma}
\def\U{\Upsilon}

\def\ca{{\cal A}}
\def\cb{{\cal B}}

\def\cm{{\cal M}}

\def\car{{\cal R}}

\def\dslash{\not{\hbox{\kern-2pt $\partial$}}}
\def\Dslash{\not{\hbox{\kern-4pt $D$}}}
\def\pslash{\not{\hbox{\kern-2.3pt $p$}}}
 \newtoks\slashfraction
 \slashfraction={.13}
 \def\slash#1{\setbox0\hbox{$ #1 $}
 \setbox0\hbox to \the\slashfraction\wd0{\hss \box0}/\box0 }
 
\font\ro=cmsy10                          
\def\kcr{{\hbox{\ro \char'170}}}                
\def\ktl{{\hbox{\ro \char'170}}}        
\def\ktr{{\hbox{\ro \char'170}}}        
\def\kbl{{\hbox{\ro \char'170}}}        
\def\kbr{{\hbox{\ro \char'170}}}        

\def\plpl{\raise-2pt\hbox{$\raise3pt\hbox{$_+$}\hskip-6.67pt\raise0.0pt
\hbox{$^+$}\hskip 0.01pt$}}
\def\mimi{\raise-2pt\hbox{$\raise3pt\hbox{$_-$}\hskip-6.67pt\raise0.0pt
\hbox{$^-$}\hskip 0.01pt$}} 

\def\bo{{\raise.15ex\hbox{\large$\Box$}}}               
\def\pa{\partial}                                       
\def\TH{{\raise.2ex\hbox{$\displaystyle \bigodot$}\mskip-4.7mu \llap H \;}}
\def\face{{\raise.2ex\hbox{$\displaystyle \bigodot$}\mskip-2.2mu \llap {$\ddot
        \smile$}}}                                      

\def\pp{{\mathchoice
          {
              \kern 1pt%
              \raise 1pt
              \vbox{\hrule width5pt height0.4pt depth0pt
                    \kern -2pt
                    \hbox{\kern 2.3pt
                          \vrule width0.4pt height6pt depth0pt
                          }
                    \kern -2pt
                    \hrule width5pt height0.4pt depth0pt}%
                    \kern 1pt
           }
            {
              \kern 1pt%
              \raise 1pt
              \vbox{\hrule width4.3pt height0.4pt depth0pt
                    \kern -1.8pt
                    \hbox{\kern 1.95pt
                          \vrule width0.4pt height5.4pt depth0pt
                          }
                    \kern -1.8pt
                    \hrule width4.3pt height0.4pt depth0pt}%
                    \kern 1pt
            }
            {
              \kern 0.5pt%
              \raise 1pt
              \vbox{\hrule width4.0pt height0.3pt depth0pt
                    \kern -1.9pt  
                    \hbox{\kern 1.85pt
                          \vrule width0.3pt height5.7pt depth0pt
                          }
                    \kern -1.9pt
                    \hrule width4.0pt height0.3pt depth0pt}%
                    \kern 0.5pt
            }
            {
              \kern 0.5pt%
              \raise 1pt
              \vbox{\hrule width3.6pt height0.3pt depth0pt
                    \kern -1.5pt
                    \hbox{\kern 1.65pt
                          \vrule width0.3pt height4.5pt depth0pt
                          }
                    \kern -1.5pt
                    \hrule width3.6pt height0.3pt depth0pt}%
                    \kern 0.5pt
            }
        }}

  \def\mm{{\mathchoice
                       {
                             \kern 1pt
               \raise 1pt    \vbox{\hrule width5pt height0.4pt depth0pt
                                  \kern 2pt
                                  \hrule width5pt height0.4pt depth0pt}
                             \kern 1pt}
                       {
                            \kern 1pt
               \raise 1pt \vbox{\hrule width4.3pt height0.4pt depth0pt
                                  \kern 1.8pt
                                  \hrule width4.3pt height0.4pt depth0pt}
                             \kern 1pt}
                       {
                            \kern 0.5pt
               \raise 1pt
                            \vbox{\hrule width4.0pt height0.3pt depth0pt
                                  \kern 1.9pt
                                  \hrule width4.0pt height0.3pt depth0pt}
                            \kern 1pt}
                       {
                           \kern 0.5pt
             \raise 1pt  \vbox{\hrule width3.6pt height0.3pt depth0pt
                                  \kern 1.5pt
                                  \hrule width3.6pt height0.3pt depth0pt}
                           \kern 0.5pt}
                       }}

\def\sp#1{{}^{#1}}                              
\def\Tilde#1{\widetilde{#1}}                    
\def\Hat#1{\widehat{#1}}                        
\def\Bar#1{\overline{#1}}                       
\def\leftrightarrowfill{$\mathsurround=0pt \mathord\leftarrow \mkern-6mu
        \cleaders\hbox{$\mkern-2mu \mathord- \mkern-2mu$}\hfill
        \mkern-6mu \mathord\rightarrow$}
\def\dvec#1{\vbox{\ialign{##\crcr
        \leftrightarrowfill\crcr\noalign{\kern-1pt\nointerlineskip}
        $\hfil\displaystyle{#1}\hfil$\crcr}}}           

\def\fracm#1#2{\hbox{\large{${\frac{{#1}}{{#2}}}$}}}
\def\frac#1#2{{\textstyle{#1\over\vphantom2\smash{\raise.20ex
        \hbox{$\scriptstyle{#2}$}}}}}                   
\def\sfrac#1#2{{\vphantom1\smash{\lower.5ex\hbox{\small$#1$}}\over
        \vphantom1\smash{\raise.4ex\hbox{\small$#2$}}}} 
\def\bfrac#1#2{{\vphantom1\smash{\lower.5ex\hbox{$#1$}}\over
        \vphantom1\smash{\raise.3ex\hbox{$#2$}}}}       
\def\afrac#1#2{{\vphantom1\smash{\lower.5ex\hbox{$#1$}}\over#2}}    

\newskip\humongous \humongous=0pt plus 1000pt minus 1000pt
\def\caja{\mathsurround=0pt}
\def\eqalign#1{\,\vcenter{\openup2\jot \caja
        \ialign{\strut \hfil$\displaystyle{##}$&$
        \displaystyle{{}##}$\hfil\crcr#1\crcr}}\,}
\newif\ifdtup

\def\ref#1{$\sp{#1)}$}

\parskip=\medskipamount                 
\thicklines                         

\def\oldheadpic{                                
        \setlength{\unitlength}{.4mm}
        \thinlines
        \par
        \begin{picture}(349,16)
        \put(325,16){\line(1,0){4}}
        \put(330,16){\line(1,0){4}}
        \put(340,16){\line(1,0){4}}
        \put(335,0){\line(1,0){4}}
        \put(340,0){\line(1,0){4}}
        \put(345,0){\line(1,0){4}}
        \put(329,0){\line(0,1){16}}
        \put(330,0){\line(0,1){16}}
        \put(339,0){\line(0,1){16}}
        \put(340,0){\line(0,1){16}}
        \put(344,0){\line(0,1){16}}
        \put(345,0){\line(0,1){16}}
        \put(329,16){\oval(8,32)[bl]}
        \put(330,16){\oval(8,32)[br]}
        \put(339,0){\oval(8,32)[tl]}
        \put(345,0){\oval(8,32)[tr]}
        \end{picture}
        \par
        \thicklines
        \vskip.2in}
\def\oldtitle#1#2#3#4{\oldheadpic\begin{center}\vglue.5in{\large\bf #1}\\[.6in]
        {#2}\\[.1in] {\it Department of Physics and Astronomy}\\
        {\it University of Maryland, College Park, MD 20742}\\[.6in]
        Physics Publication \#{#3}\\ {#4}\\[1.5in] {\bf ABSTRACT}\\[.1in]
        \end{center} \begin{quotation}}                 
\def\oldTitle#1#2#3#4#5#6#7{\oldheadpic\begin{center} \vglue .4in
        {\large\bf #1}\\[.4in]
        {#2}\\[.1in] {\it Department of Physics and Astronomy}\\
        {\it University of Maryland, College Park, MD 20742}\\[.1in]
        {#3}\\[.1in] {\it {#4}}\\ {\it {#5}}\\[.4in]
        Physics Publication \#{#6}\\ {#7}\\[.5in] {\bf ABSTRACT}\\[.1in]
        \end{center} \begin{quotation}}                 
\def\border{                                            
        \setlength{\unitlength}{1mm}
        \newcount\xco
        \newcount\yco
        \xco=-21
        \yco=12
        \begin{picture}(140,0)
        \put(\xco,\yco){$\ktl$}
        \advance\yco by-1
        {\loop
        \put(\xco,\yco){$\kcr$}
        \advance\yco by-2
        \ifnum\yco>-240
        \repeat
        \put(\xco,\yco){$\kbl$}}
        \xco=158
        \yco=12
        \put(\xco,\yco){$\ktr$}
        \advance\yco by-1
        {\loop
        \put(\xco,\yco){$\kcr$}
        \advance\yco by-2
        \ifnum\yco>-240
        \repeat
        \put(\xco,\yco){$\kbr$}}
        \put(-20,13){\tiny University of Maryland Elementary Particle
Physics University of Maryland Elementary Particle Physics University of
Maryland Elementary Particle Physics}
        \put(-20,-241.5){\tiny University of Maryland Elementary
Particle Physics University of Maryland Elementary Particle Physics
University of Maryland Elementary Particle Physics}
        \end{picture}
        \par\vskip-8mm}
\def\bordero{                                           
        \setlength{\unitlength}{1mm}
        \newcount\xco
        \newcount\yco
        \xco=-31
        \yco=12
        \begin{picture}(140,0)
        \put(\xco,\yco){$\ktl$}
        \advance\yco by-1
        {\loop
        \put(\xco,\yco){$\kclr}
        \advance\yco by-2
        \ifnum\yco>-240
        \repeat
        \put(\xco,\yco){$\kbl$}}
        \xco=151
        \yco=12
        \put(\xco,\yco){$\ktr$}
        \advance\yco by-1
        {\loop
        \put(\xco,\yco){$\kcr$}
        \advance\yco by-2
        \ifnum\yco>-240
        \repeat
        \put(\xco,\yco){$\kbr$}}
        \put(-20,12){\ooo bacdefghidfghghdhededbihdgdfdfhhdheidhdhebaaahjhhdahba

hgdedge
   hgfdiehhgdigicba}
        \put(-20,-241.5){\ooo ababaighefdbfghgeahgdfgafagihdidihiidhiagfedhadbfd

ecdcdfa
   gdcbhaddhbgfchbgfdacfediacbabab}
        \end{picture}
        \par\vskip-8mm}
\def\headpic{                                           
        \indent
        \setlength{\unitlength}{.4mm}
        \thinlines
        \par
        \begin{picture}(29,16)
        \put(165,16){\line(1,0){4}}
        \put(170,16){\line(1,0){4}}
        \put(180,16){\line(1,0){4}}
        \put(175,0){\line(1,0){4}}
        \put(180,0){\line(1,0){4}}
        \put(185,0){\line(1,0){4}}
        \put(169,0){\line(0,1){16}}
        \put(170,0){\line(0,1){16}}
        \put(179,0){\line(0,1){16}}
        \put(180,0){\line(0,1){16}}
        \put(184,0){\line(0,1){16}}
        \put(185,0){\line(0,1){16}}
        \put(169,16){\oval(8,32)[bl]}
        \put(170,16){\oval(8,32)[br]}
        \put(179,0){\oval(8,32)[tl]}
        \put(185,0){\oval(8,32)[tr]}
        \end{picture}
        \par\vskip-6.5mm
        \thicklines}
\def\title#1#2#3#4{\border\headpic {\hbox to\hsize{#4 \hfill UMDEPP #3}}\par
        \begin{center} \vglue .5in {\large\bf #1}\\[.6in]
        {#2}\\[.1in] {\it Department of Physics and Astronomy}\\
        {\it University of Maryland, College Park, MD 20742}\\[1.5in]
        {\bf ABSTRACT}\\[.1in] \end{center} \begin{quotation}}  
\def\Title#1#2#3#4#5#6#7{\border\headpic
        {\hbox to\hsize{#7 \hfill UMDEPP #6}}\par
        \begin{center} \vglue .4in {\large\bf #1}\\[.4in]
        {#2}\\[.1in] {\it Department of Physics and Astronomy}\\
        {\it University of Maryland, College Park, MD 20742}\\[.1in]
        {#3}\\[.1in] {\it {#4}}\\ {\it {#5}}\\[.5in] {\bf ABSTRACT}\\[.1in]
        \end{center} \begin{quotation}}                 
\def\endtitle{\end{quotation}\newpage}                  

\def\dpx{{{ =\hskip-3.75pt{\lin}}\hskip3.75pt }}

\def\qd{{\kern0.5pt
                   q \kern-5.05pt \raise5.8pt\hbox{$\textstyle.$}\kern
0.5pt}}

\global\newcount\secno \global\secno=0
\def\newsec#1{\newpage \global\advance\secno by1\message{[#1]}
              \section{#1}}

\def\fracm#1#2{\hbox{\large{${\frac{{#1}}{{#2}}}$}}}

\def\lin{\vrule width0.5pt height5pt depth1pt}

\def\pa{\partial}                                       
\def\Bar#1{\overline{#1}}                               

\def\cdp{{\cal D}_{\pp}}
\def\cdm{{\cal D}_{\mm}}

\def\ep{e_{\pp}}
\def\em{e_{\mm}}
\def\op{\omega_{\pp}}
\def\om{\omega_{\mm}}
\def\cm{{\cal M}}
\def\cap#1#2{{\cal A}_{\pp{#1}}{}^{#2}}
\def\cam#1#2{{\cal A}_{\mm{#1}}{}^{#2}}
\def\CY#1#2{{\cal Y}_{#1}{}^{#2}}

\def\sp#1{\psi_{\pp}{}^{+{#1}}}
\def\spb#1{{\overline\psi}_{\pp}{}^{+}{}_{#1}}
\def\sm#1{\psi_{\mm}{}^{+{#1}}}
\def\smb#1{{\overline\psi}_{\mm}{}^{+}{}_{#1}}

\def\sig#1{\Sigma^{+{#1}}}
\def\sigb#1{{\overline\Sigma}^{+}{}_{#1}}
\def\car{{\cal R}}
\def\CF#1#2{{\cal F}_{#1}{}^{#2}}

\def\del#1#2{\delta_{#1}{}^{#2}}

\def\ca{{\cal A}}
\def\cab{{\overline{\cal A}}}
\def\cb{{\cal B}}
\def\cbb{{\overline{\cal B}}}
\def\si#1{\psi^{-{#1}}}
\def\sib#1{{\overline{\psi}}^{-}{}_{#1}}

\def\f{\phi}
\def\fai#1#2{\phi_{#1}{}^{#2}}
\def\lm#1{\lambda^{-}{}_{#1}}
\def\lmb#1{{\overline\lambda}^{-}{}^{#1}}

\def\ai#1{{\cal A}_{#1}}
\def\aib#1{{\overline{\cal A}}^{#1}}
\def\pm{\pi^{-}}
\def\pmb{{\overline\pi}^{-}}

\begin{document}

\def\gfrac#1#2{\frac {\scriptstyle{#1}}
        {\mbox{\raisebox{-.6ex}{$\scriptstyle{#2}$}}}}
\def\gg{{\hbox{\sc g}}}
\border\headpic {\hbox to\hsize{June 1996 \hfill {UMDEPP 96-110}}}
\par
\setlength{\oddsidemargin}{0.3in}
\setlength{\evensidemargin}{-0.3in}
\begin{center}
\vglue .08in
{\large\bf A Canticle on \\
(4,0) Supergravity-Scalar Multiplet Systems\\
for a ``Cognoscente''
\footnote {Supported in part by National 
Science Foundation Grant PHY-91-19746 \newline ${~~~~~}$ and by NATO 
Grant CRG-93-0789}  }
\\[.72in]
R. Dhanawittayapol, S. James Gates, Jr. \\
and Lubna Rana
\\[0.5in]
{\it Department of Physics\\ 
University of Maryland\\ 
College Park, MD 20742-4111  USA}\\[.2in] 
{\bf {\tt gates@umdhep.umd.edu}}\\
{\bf {\tt lubna@umdhep.umd.edu}}\\[1.8in]

{\bf ABSTRACT}\\[.002in]
\end{center}
\begin{quotation}
{Extending prior investigations, we study three of the the four 
{\it {distinct}} minimal (4,0) scalar multiplets coupled to (4,0) 
supergravity. It is found that the scalar multiplets manifest their 
differences at the component level by possessing totally different 
couplings to the supergravity fields. Only the SM-I multiplet possesses 
a conformal coupling.  For the remaining multiplets, terms linear in 
the world sheet curvature and/or SU(2) gauge field strengths are 
required to appear in the action by local supersymmetry.}  

\endtitle

\noindent
{\bf {(I.) Introduction}}

One of the marvelous ways in which science differs from most human
endeavors is that it is self-correcting.  Furthermore, there are
right answers and there are wrong answers.  As such, the beliefs of 
even experts can be changed upon proof that their misconceptions are
not grounded in reality.  Some might say that parts of theoretical physics 
are not as well grounded in reality.  However, even here we have rules 
of mathematical and logical consistency that act as a veto to the
long term support of misconceptions and falsehoods.

Some time ago we argued that 2D representations of extended 
supersymmetry likely possess inherent ambiguities that permit the 
existence of many more distinct representations than one might
naively guess. In the case of (4,0) supersymmetry we showed that
this was precisely the case \cite{A}. Remarkably enough if one
considers the simplest (4,0) supersymmetric representation, the
minimal scalar multiplet, it appears in {\underline {four}} different
varieties. Our observation acted as a generalization to that made 
by Witten in his work on ADHM non-linear $\s$-models \cite{B} where 
it was proposed that two such multiplets (a (4,0) scalar multiplet 
and its twisted version) exist.

The work of \cite{A} was partially inspired by Witten's remarkable
proposal.  Prior to the work on the ADHM $\s$-model, the only
``twisting'' known was related to 2D parity.  In the language of
conformal field theory, ``twisting'' is equivalent to the statement
that the holomorphic and anti-holomorphic parts of a theory
are not quite identical.  However, in a (4,0) theory, there is 
{\underline {no}} anti-holomorphic part of the theory to consider.
So the question arose, ``What is being twisted in Witten's ADHM 
models?''  The answer provided by the work in \cite{A} is that 
the twists in the ADHM $\s$-model take place with respect to
the SU(2) group of conformal (4,0) supergravity.

Our (4,0) results obviously have implications for 2D (4,4) or $N$ = 4
systems since (4,0) theories can be embedded into (4,4) theories.
These implications were investigated in a study of 2D (4,4) 
hypermultiplets \cite{C} that was completed some time ago. 
The result of that study was that {\underline {eight}} distinct 2D, 
$N$ = 4 hypermultiplets were found!  This results goes against the 
general belief of some ``experts'' who have expressed unreasonable 
and unfounded skepticism. The criticism of our $N$ = 4 results has been 
based on the naive statement that all of our ``... distinct (4,0) 
scalar multiplet theories must be related by field redefinitions.'' Our 
response to this has been that the multiplets are not related by field 
re-definitions but are related by automorphisms of the supersymmetry 
parameters. The existence of such automorphisms has been known for 
over a decade.  This is precisely the relation between 2D, $N$ = 2 
chiral multiplets and 2D, $N$ = 2 twisted chiral multiplets \cite{D}.

When we describe two multiplets as being distinct or inequivalent,
we mean that the set of all possible dynamics that can be described 
by use of one of the multiplets is distinct from that that can be 
described by use of the other one of the multiplets or by a linear 
combination of the two.  The case of the 2D, $N$ = 2 chiral multiplets 
versus 2D, $N$ = 2 twisted chiral multiplets is a prototype example of 
this statement.  The class of non-linear $\s$-models using 2D, $N$ = 2 
chiral multiplets must necessarily possess a K\" ahler geometry. Since 
a K\" ahler geometry is Riemannian, it has no torsion.  On the other 
hand, the class of non-linear $\s$-models using 2D, $N$ = 2 chiral 
multiplets {\underline {and}} 2D, $N$ = 2 twisted chiral multiplets can 
describe a complex geometry with torsion. Therefore, 2D, $N$ = 2 chiral 
multiplets are distinct from 2D, $N$ = 2 twisted chiral multiplets.

The simplest and most direct way to show that our (4,0) scalar
multiplets are distinct is to explicitly demonstrate how the
dynamics of the multiplets differ. There are many ways to do this.
For example in \cite{C} we showed how the massive dynamics of
$N$ = 4 hypermultiplets differ. One other demonstration of the
different nature of the (4,0) hypermultiplets is to couple 
them to (4,0) supergravity.  This will be the topic pursued
in the present work. \newline

There is a sense (explained in ref. \cite{C}) in which all of the 
multiplets are ``twisted'' versions of one another.  For example,
SM-I and SM-III can be obtained one from the other by simply switching 
the Grassmann parity of all of the fields within a multiplet.  We call
this the ``Klein flip.''

\noindent
{\bf {(II.) The Varieties of Minimal (4,0) Scalar Multiplet Theory}}

In a previous work we have pointed out a generally unrecognized fact 
regarding (4,0) minimal irreducible scalar multiplet theories. Namely 
there are four distinct such theories.   We denote these theories by
SM-I, SM-II, SM-III and SM-IV.  The field content of these are summarized 
in the 
following table.
\begin{center}
\renewcommand\arraystretch{1.2}
\begin{tabular}{|c|c| }\hline
${\rm Multiplet}$  & ${\rm Field~Content}$  \\ \hline 
\hline
$~~~{\rm SM-I}~~~$ &  $~~({\cal A},~{\cal B}, ~ \psi^{- i} )~~$ \\ \hline
$~~~{\rm SM-II}~~~$ &  $~~(\phi,~\phi_i {}^j , ~ \l^- {}_i )~~$   \\ \hline
$~~~{\rm SM-III}~~~$ & $ ({\cal A}_i,~\rho^- , ~ \pi^-  )$   \\ \hline
$~~~{\rm SM-IV}~~~$ &  $ ({\cal B}_i,~\psi^- , ~ \psi^-_i {}^j  ) $   \\ \hline
\end{tabular}
\end{center}
\vskip.2in
\centerline{{\bf Table I}} 
In this table each Latin letter index appended to a field denotes the defining 
representation of SU(2). All fields with two such indices are traceless.  Each 
multiplet contains four bosons and four fermions. The bosons are ${\cal A},\, 
{\cal B}, \, \phi, \, \phi_i {}^j , \, {\cal A}_i $ and $ {\cal B}_i $ and of 
these only $\phi$ and $\phi_i {}^j$ are real.  Similarly, the fermions 
$\psi^-$ and $\psi^-_i {}^j$ are real (Majorana). 
\newline

\noindent
{\bf {(III.) Prepotential Formulation of Minimal (4,0) Scalar Multiplet Theory}}

The starting point of the manifestly supersymmetric quantization of a classical
theory possessing supersymmetry is the construction of the description of that 
theory in terms of unconstrained superfields called pre-potentials. The main
advantage of such a formulation is that it allows the powerful supergraph
technique to be utilized in the exploration of quantum behavior of the
classical theory. The most striking outcomes of such an approach are the
derivation of non-renormalization theorems.  
 
The (4,0) SM-I (scalar multiplet one) theory described in terms of constrained
superfields is given by, 
$$ \eqalign{ D_{+ i} {\cal A} ~ = ~ & 2 C_{ij} \psi^- 
             {}^{j} ~~~~~, ~~~~~~~ {\Bar D}_+ {}^i {\cal A} ~ =~ 0 ~~~, \cr
             {\Bar D}_+ {}^i {\cal B} ~ =~ & i 2  \psi^- 
             {}^{i} ~~~~~~~~, ~~~~~~~\,  D_{+ i} {\cal B} ~ =~ 0 ~~~, \cr
              {\Bar D}_+ {}^i \psi^- {}^{j} ~ =~ & i C^{ij} 
             \pa_{\dpx} {\cal A} ~~~~,~~~~\, D_{+ i} \psi^- {}^{j} ~=~
             \d_i {}^j  \pa_{\dpx} {\cal B}     ~~~. } 
\eqno(1) $$
These superdifferential constraints can be solved explicitly in terms of a 
prepotential superfield ($P_{\mm - } {}^i$) that is subject to {\underline {no}} 
differential constraints.
$$ {\cal A} ~\equiv ~ {\Bar D}_{\pp}^{~2} D_{+ i} P_{\mm - } {}^i ~~~~,~~~~ 
{\cal B} ~\equiv ~ - i C_{i j} D_{\pp}^2 {\Bar D}_+ {}^i P_{\mm - } {}^j ~~~, $$
$$  \psi^- {}^i  ~\equiv ~ - \frac 12 C^{i j} D_{+ j} {\Bar D}_{\pp}^{~ 2} D_{+ k} 
P_{\mm - } {}^k ~~~~,  \eqno(2) $$
where $D_{+ i} D_{+ j} \equiv C_{i j} D_{\pp}^2$ and ${\Bar D}_+ {}^i
{\Bar D}_+ {}^j \equiv C^{i j} {\Bar D}_{\pp}^{~2}$.
Using the algebra of the supercovariant derivatives it can be shown that
the results in (1) follow now as simple consequences. The prepotential
superfield is actually a gauge superfield. The quantities ${\cal A}$, 
${\cal B}$ and $\psi^- {}^i$ are invariant under the gauge transformation 
given by 
$$ \d_G P_{\mm - } {}^i ~= ~ D_{+ j} \L^{(i j)}_{\mm \mm} ~+~ 
{\Bar D}_+ {}^j {\Hat \L}_{\mm \mm j} {}^i ~~~~,~~~~ {\Hat \L}_{\mm \mm i} 
{}^i ~=~ 0 ~~~~. 
\eqno(3) $$

The (4,0) SM-II (scalar multiplet two) theory described in terms of constrained
superfields is given by, 
$$ \eqalign{ D_{+ i}~ \phi ~ =~ & ~i  \lambda^- {}_i  ~~~, ~~~ \phi ~=~ 
\phi^*  ~~~~, {~~~~~~~~~~} \cr
{~~~~~~~~~~~~} D_{+ i} ~  \phi_{j}{}^{k}~ =~&~ 2 \d_i {}^{k} \lambda^- {}_{j} - 
 \d_{j} {}^{k} \lambda^- {}_{i}  ~~~, ~~~
\phi_{i}{}^{j} ~=~ (\phi_{j}{}^{i})^* ~~~, ~~~
\phi_{i}{}^{i} ~=~ 0 ~~~, \cr
{\Bar D}_+ {}^i ~ \lambda^- {}_j  ~ =~ & \d_j {}^i \pa_{\dpx} 
\phi ~+~ i  \pa_{\dpx} \phi_j{}^i ~~~, ~~~ D_{+ i} ~ \lambda^- {}_j  ~ =~ 0
~~~. } \eqno(4) $$
These superdifferential constraints can be solved explicitly in terms of a 
prepotential superfield ($\Psi_{- ~j} $) that is subject to a chirality
constraint ${\Bar D}_+ {}^i \Psi_{- ~j} = 0 $,
$$ \phi ~\equiv ~ - i [~  C^{i j} D_{+ i} \Psi_{- ~j} ~+~ C_{i j} {\Bar D}_+ {}^i
{\Bar \Psi}_{-}{}^{~j} ~] ~~~~, {~~~~~~~~~~~~~~~~~~~~}
{~~~~~~~~~~~~~~~~~~~~~~~~~~}$$ 
$$\phi_{i}{}^{j}  ~\equiv ~ 2 [~  C^{j k} D_{+ k} \Psi_{- ~i} ~-~ C_{i k} {\Bar 
D}_+ {}^j {\Bar \Psi}_{-}{}^{~k} ~] ~-~ \d_i {}^j [~  C^{k l} D_{+ k} \Psi_{- ~l} 
~-~ C_{k l} {\Bar D}_+ {}^k {\Bar \Psi}_{-}{}^{~l} ~] ~~~, $$
$$ \lambda^- {}_{i} ~\equiv ~  D_{\pp}^2 \Psi_{- ~i} ~-~ i 2 C_{i j} \pa_{\dpx}
{\Bar \Psi}_{-} {}^{ ~j} ~~~. {~~~~~~~~~~~~~~~~~~~~} {~~~~~~~~~~~~~~}
{~~~~~~~~} {~~~~~~~~} {~~~~~~~~}\eqno(5) $$
The field strength superfields above are invariant under the gauge transformation,
$$\d_G  \Psi_{- ~i} ~=~ {\Bar D}_{\pp}^{~ 2}[~ D_{+ i} \L_{\mm \mm} ~+~ i
D_{+ j} {\Tilde \L}_{\mm \mm i} {}^j ~]  ~~~~,~~~~ {\Tilde \L}_{\mm \mm i} 
{}^i ~=~ 0 ~~~~, $$
$$ \L_{\mm \mm} ~=~ (\L_{\mm \mm})^* ~~~~,~~~~ {\Tilde \L}_{\mm \mm i} {}^j
~=~ ({\Tilde \L}_{\mm \mm j} {}^i)^* ~~~~. {~~~~~~~~~~~~~~~~~~~} \eqno(6) $$ 

The (4,0) SM-III (scalar multiplet three) theory described in terms of constrained
superfields is given by, 
$$ \eqalign{ 
D_{+ i} {\cal A}_j ~ =~ &  C_{ij} \pi^-  ~~~~~~~~~~~,~~~~ {\Bar D}_+ {}^i {\cal A}_j 
~=~ \d_j {}^i \rho^-  ~~~, \cr
D_{+ i} \rho^- ~ =~ &  i 2 \,  \pa_{\dpx} {\cal A}_i ~~~~~~~~~, ~~~~{\Bar D}_+ {}^i
\rho^- ~ =~ 0 ~~~~,   \cr
{\Bar D}_+ {}^i \pi^- ~ =~ & i2 \, C^{ij} \,  \, \pa_{\dpx}  
{\cal A}_j ~~~~,  ~~~\, D_{+ i} \pi^- ~ =~ 0 ~~~~. } 
\eqno(7)$$
The solution to this set of superdifferential equations can be expressed in terms
of two independent prepotential superfields ($\S_-$ and $\U_-$) that each satisfies 
a chirality constraint (${\Bar D}_+ {}^i \S_- = 0 $ and ${\Bar D}_+ {}^i \U_- = 0$). 
The explicit form of this solution is,
$${\cal A}_i ~ =~  D_{+ i} \S_- ~+~ C_{i j} {\Bar D}_+ {}^j {\Bar \U}_- ~~~~, $$ 
$$\pi^- ~ =~ D_{\pp}^2 \S_- ~-~ i 2 \, \pa_{\dpx} {\Bar \U}_- ~~~~, $$ 
$$\rho^- ~ =~ {\Bar D}_{\pp}^{~ 2} {\Bar \U}_- ~+~ i 2 \, \pa_{\dpx}  \S_-
~~~~.  \eqno(8)$$ 
These are invariant under the following gauge transformation,
$$\d_G \S_- ~=~ {\Bar D}_{\pp}^{~ 2} D_{+ i} \L_{\mm \mm} {}^i  ~~~~,~~~~
\d_G \U_- ~=~ -\,  C^{i j} {\Bar D}_{\pp}^{~ 2} D_{+ i} \L_{\mm \mm ~j}^*  ~~~~. 
\eqno(9)$$

The (4,0) SM-IV (scalar multiplet four) theory described in terms of constrained
superfields is given by, 
$$ \eqalign{ {~~~~~}
{\Bar D}_+ {}^i {\cal B}_{j}~ =~ & \d_j {}^i \, \psi^- ~+~ i 2 \, \psi^- {}_j 
{}^{i} ~~~~,~~~~\, {~~~~~} D_{+ i} {\cal B}_{j}~ =~ 0 ~~~~,  \cr
D_{+ i} \psi^-~ =~ &  i  \, \pa_{\dpx} {\cal B}_{i}   ~~~~~~,{~~~~~~~~~~~~~~~~}
{~~~~~~~~~~} \psi^- 
~=~ (\psi^-)^* ~~~~, \cr
D_{+ i} \psi^- {}_j {}^{k}~ =~ &  \, \d_i {}^k \pa_{\dpx} 
{\cal B}_{j} - \frac12 \d_j {}^{k} \,  \pa_{\dpx} {\cal
 B}_{i} ~~~~,~~~~ {~~~}\psi^- {}_i {}^{j} ~=~ (\psi^- {}_j 
{}^i)^* ~~~~. } 
\eqno(10) $$
As above, these superdifferential constraints have an explicit solution
given in terms of two independent prepotential superfields $U_{\mm -}$ and 
$V_{\mm - i}{}^j$. In order to write the solution to the constraints, it is 
convenient to define $S_{\mm -}$ and $T_{\mm - ~i}{}^j$ as $S_{\mm -} 
\equiv U_{\mm -} + (U_{\mm -} )^*$ and $T_{\mm -  ~i}{}^j \equiv V_{\mm - 
i}{}^j +  (V_{\mm - i}{}^j )^*$ so that $S_{\mm -} = - (S_{\mm -})^*$ and $
 T_{\mm - ~i}{}^j = - ( T_{\mm - ~j}{}^i )^* $) are subject to {\underline 
{no}} differential constraints.
$${\cal B}_{i} ~\equiv~  D_{\pp}^2 [~ i C_{i j} {\Bar D}_+ {}^j S_{\mm -}
~+~ C_{j k} {\Bar D}_+ {}^j T_{\mm - ~i}{}^k ~] ~~~~,{~~~~~~~~~~~} {~~~~~~~~~~}$$
$$\psi^- ~\equiv~ i \frac 12 [~ C_{i j} {\Bar D}_+ {}^i D_{\pp}^2 {\Bar D}_+ {}^j
S_{\mm -} ~+~ 2 \, \pa_{\dpx} D_{+ i} {\Bar D}_+ {}^j T_{\mm - ~j}{}^i ~] ~~~~, 
{~~~~~~~~~} $$
$$\psi^- {}_i {}^j~ \equiv~ \frac 14 [~ ( C_{i k} {\Bar D}_+ {}^j D_{\pp}^2 
{\Bar D}_+ {}^k - C^{j k} D_{+ i} {\Bar D}_{\pp}^{~ 2} D_{+ k}
) S_{\mm -} ~] {~~~~~~~~~~~~~~~~~~~} $$
$$ {~~~~~~~~~~~~~}- i \frac 14 [~  C_{k l} {\Bar D}_+ {}^j D_{\pp}^2 
{\Bar D}_+ {}^k T_{\mm - ~i}{}^l + C^{k l} D_{+ i} {\Bar D}_{\pp}^{~ 2} D_{+ k}
T_{\mm - ~l}\,{}^j ~]  ~~~~. \eqno(11) $$
These are invariant under a set of gauge variations given by,
$$\d_G U_{\mm -} ~=~  D_{+ i} {\Hat \L}_{\mm \mm} \, {}^i
 ~~~~, {~~~~~~~~~~~~~~} $$
$$\d_G V_{\mm - ~i}{}^j ~=~ - i \frac 23 \,D_{+ i} {\Hat \L}_{\mm \mm} \, 
{}^j   ~+~ i \frac 13  \d_i {}^j \, D_{+ k} {\Hat \L}_{\mm \mm} \, {}^k 
~+~ C_{i p} D_{+ q} {\Hat \L}_{\mm \mm} \, {}^{ (j p q)} 
  ~~~~. \eqno(12) $$

This completes the unconstrained superfield description of the various scalar
multiplets with manifest (4,0) supersymmetry. As can be seen, each of the 
scalar multiplets is described by a (set of) gauge prepotential superfields.
These provide the fundamental superfields that can (in principle) be quantized
and used to generate supergraph rules. One other interesting observation is
that the 2D Lorentz representation of the gauge parameter superfield for
{\underline {all}} the scalar multiplets is the same. All such superfields
transform as the minus two representation of the 2D lorentz group.

For the SM-I, SM-II, SM-III and SM-IV models, the free actions are obtained 
from the following respective superspace expressions,

$$ {\cal S}_{\rm {SM-I}} ~=~ \Big[ ~\int d^2 \s \, d^2 \z^{\pp} ~ [ ~  - i
\frac{1}{4}\cbb\pa_{\mm}\ca ~]  ~+~   {\rm h.} {\rm c.} ~ 
\Big] ~~~, 
$$
$$ {\cal S}_{\rm {SM-II}} ~=~ \Big[ ~\int d^2 \s \, d^2 \z^{\pp} ~  [~-\,  
\frac 12  \Psi_{- i} \pa_{\mm} {\Bar \l}^{- i} ~] ~+~   {\rm h.} {\rm c.} ~ 
\Big] ~~~~, 
$$
$$ {\cal S}_{\rm {SM-III}} ~=~ \Big[ ~\int d^2 \s \, d^2 \z^{\pp}~ [~ - \, 
\frac 18 \S_- \pa_{\mm} {\Bar \pi}^-  ~+~  \frac 18 \U_- \pa_{\mm} {\rho}^- 
~]  ~+~ 
{\rm h.} {\rm c.} ~ \Big] ~~~~, 
$$
$$ {\cal S}_{\rm {SM-IV}} ~=~ \Big[ ~\int d^2 \s \, d^2 \z^{\pp} ~ [ ~  - i
\frac{1}{4} C_{i j} \cbb^i \pa_{\mm}\cbb^j ~]  ~+~   {\rm h.} {\rm c.} ~ 
\Big] ~~~. 
\eqno(13)
$$
Now the critical feature about these expressions is that in order to write the
actions for SM-II and SM-III, we had to explicitly express them in terms of 
prepotentials.  This is vastly different from the SM-I and SM-IV theory where 
their chiral actions were totally expressible {\it {solely}} in terms of the 
field strength superfields.  We know that the actions for SM-II and SM-III above 
only involve the component fields contained in the field strength superfields 
because these actions are {\it {gauge}} {\it {invariant}} with respect to the 
prepotential gauge transformations.

\noindent
{\bf { (IV.) (4,0) Supergravity Theory and Superstrings}}

The supergeometry of (p, 0) supergravity has been known for some time \cite{O}. It is
simple to specialize to the case of p = 4.  There is a supergravity covariant 
derivatives 
$\nabla_A \equiv ( \nabla_{+ i}, \, 
{\Bar \nabla}_+ {}^i , \, \nabla_{\dpx} , \, \nabla_= )$ that can be expanded over
a supervielbein ($E_A {}^M$), Lorentz spin-connection ($\o_A$) and SU(2)
gauge connection (${\cal A}_{A \, i} {}^j$),
$$
\nabla_A ~=~ E_A {}^M {\rm D}_M ~+~ \o_A {\cal M} ~+~ i \, {\cal A}_{A \, i} {}^j
{\cal Y}_j {}^i ~~~~.  \eqno(14)
$$
Above ${\rm D}_M$ denotes the flat space fermi and bose derivatives ${\rm D}_M
\equiv (  {\Bar {\rm D}}_{+ i}, \, {\rm D}_+ {}^i , \, \pa_{\dpx} , \, \pa_= $). 
Similarly, ${\cal M}$ and ${\cal Y}_i {}^j$ denotes the Lorentz and SU(2) 
generators respectively.  These act on $\nabla_A$ as 
$$ 
[\, {\cal M} , \nabla_{+ i} \,] ~=~ \frac 12 \, \nabla_{+ i} ~~~~,~~~~
[\, {\cal M} , {\Bar \nabla}_+ {}^i \,] ~=~ \frac 12 \, {\Bar \nabla}_+ {}^i
 ~~~~,~~~~ [\, {\cal M} , \nabla_{\dpx} \, ] ~=~  \nabla_{\dpx}
~~~~. \eqno(15)
$$
$$
[\, {\cal Y}_j {}^k , \nabla_{+ i} \,] ~=~ \d_i {}^k \nabla_{+ j} ~-~
\frac 12 \, \d_j {}^k \nabla_{+ i} ~~~~,~~~~
[\, {\cal Y}_j {}^k , {\Bar \nabla}_+ {}^i \,] ~=~ - \d_j {}^i {\Bar \nabla}_+ 
{}^k  ~+~  \frac 12 \, \d_j {}^k {\Bar \nabla}_+ {}^i
 ~~~~,
$$
$$  [\, {\cal M} , \nabla_{=} \, ] ~=~ -\, \nabla_{=} ~~~~, ~~~~ [\, {\cal Y}_j 
{}^k , \nabla_{\dpx} \, ] ~=~ 0 ~~~~,~~~~ [\, {\cal Y}_j {}^k , \nabla_{\mm} 
\, ] ~=~ 0   ~~~~. $$
The covariant derivatives have a commutator algebra that takes the form
$$[~ \nabla_{+ i} \,, \, \nabla_{+ j} ~\} ~=~ 0 ~~~~,~~~
[~ \nabla_{+ i} \,, \, {\Bar \nabla}_+ {}^j ~\} ~=~ i 2 \d_i {}^j \, \nabla_{\dpx}
~~~~, ~~~~ [~ \nabla_{+ i} \,, \, \nabla_{\dpx} ~\} ~=~ 0 ~~~~, $$
$$ 
[~ \nabla_{+ i} \,, \, \nabla_{\mm} ~\} ~=~ - i \, [~ {\Bar \S}^+ {}_i {\cal M} 
~-~  {\Bar \S}^+{}_j {\cal Y}_i {}^j ~] ~~~~,$$
$$[~  \nabla_{\dpx} , \, \nabla_{\mm}~ \} ~=~ - \frac 12 \, [~ {\S}^{+ i} \, 
\nabla_{+ i} ~+~{\Bar \S}^+ {}_i \, {\Bar \nabla}_+ {}^i  ~+~ {\cal R} {\cal M} 
~+~ i {\cal F}_i {}^j {\cal Y}_j {}^i ~ ] ~~~~. \eqno(16) $$
These lead to a set of Bianchi identities that are solved if 
$$ {\Bar \nabla}_+ {}^i \, \S^{+ j} ~=~ 0 ~~~~,~~~~ {\nabla}_{+ i} \, \S^{+ j} ~=~
\frac 12 \, \d_i {}^j {\cal R} ~+~ i \, {\cal F}_i {}^j ~~~~, $$
$$ {\nabla}_{+ i} \, {\cal R} ~=~ i \, 2  \nabla_{\dpx} {\Bar \S}^+ {}_i ~~~~,~~~~
{\nabla}_{+ i} \, {\cal F}_j {}^k ~=~ - \, 2 \d_i {}^k \nabla_{\dpx} {\Bar \S}^+ {}_j
~+~ \d_j {}^k \nabla_{\dpx} {\Bar \S}^+ {}_i ~~~~.  \eqno(17) $$

The component fields of the (4,0) supergravity multiplet are $e_a {}^m$ (a 
zweibein), $\psi_a {}^{+ \, i}$ (SU(2) doublet gravitini) and $A_{a \, i} {}^j$ 
(gauge SU(2) triplet of auxiliary fields). The supersymmetry variations of
these may be chosen to take the forms,
$$ \d_Q e_{\pp} {}^m ~=~ \d_Q A_{\pp \, i} {}^j ~=~ 0 ~~~,~~~
\d_Q \psi_{m} {}^{+ \, i} ~=~ {\cal D}_m \e^{+ \, i} ~~~, $$
$$ \d_Q e_{\mm} {}^m ~=~ - i 2  g^{ m n} [\, {\Bar \e}^+ {}_i \psi_n {}^{+ \, i}
~+~  \e^{+ \, i} {\Bar \psi}_n {}^+ {}_i \,]  ~~~,
$$
$$ \d_Q  A_{\mm \, i} {}^j ~=~ \Big[  2 ~[ \,  \e^{+ \, j} {\Bar \psi}_{\pp , \, 
\mm} {}^+ {}_i ~-~ \frac 12 \d_i {}^j   \e^{+ \, k} {\Bar \psi}_{\pp , \, \mm} 
{}^+ {}_k \, ] ~+~  {\rm {h.\, c.}} ~  \Big]
~~~, \eqno(18) $$
where $ g^{m n} \equiv [\, e_{\pp}{}^n e_{\mm}{}^m ~+~  e_{\pp}{}^n 
e_{\mm}{}^m \,] $.

We next turn to the problem of finding (4,0) locally supersymmetric actions.
If ${\cal L}_{\mm }$ is a chiral Lagrangian ($ {\Bar \nabla_+}^i
{\cal L}_{\mm } = 0$) and ${\Bar {\cal L}}_{\mm }$ is an anti-chiral Lagrangian
($\nabla_{+ i} {\Bar {\cal L}}_{\mm } = 0 $), then component actions are 
derivable from
$$\eqalign{ \int d^2 \s d^{2}  \zeta^{\pp} ~{\cal E}^{-1} {\cal L}_{\mm } 
\mid~~~&\equiv~~~ i \int d^2 \s \left[ ~{\frac12} \,e^{-1} \,  C^{ij} \,  \left( \,
\nabla_{+i} + i 4 \, e \, {\Bar{\psi}}_{\dpx}{}^{+}{}_i \right) \right] \nabla_{+j} 
{\cal L}_{\mm }  \mid  ~~~, \cr
\int d^2 \s d^{2} \, {\Bar{\zeta}}^{\pp} ~ {\Bar{\cal E}}^{-1} \, {\Bar{\cal L}}_{
\mm } \mid~~~& \equiv~~~ i \int d^2 \s \left[ ~ {\frac12} \,e^{-1} \, C_{ij} \,  \left(
\, {\Bar \nabla}_+ {}^{i} + i 4 \, e \, \psi_{\dpx} {}^{+ i} \right) \right] {\Bar
\nabla}_+ {}^{j} {\Bar{\cal L}}_{\mm } \mid ~~~. } \eqno(19)$$
For a general Lagrangian ${\cal L}_{\mm \mm}$, the component action follows 
from
$$\eqalign{ \int d^2 \s \, d^{2} \zeta^{\pp} \, d^{2} {\Bar{\zeta}}^{\pp} ~ E^{-1} 
{\cal L}_{\mm \mm}
 ~~~&\equiv~~~\fracm 12 \int d^2 \s \, d^{2} \zeta^{\pp} \, {\cal E}^{-1} \left[ \, 
{\frac12} C_{ij} \, {\Bar \nabla}_+ {}^{i} \, {\Bar \nabla}_+ {}^{j} \,  \right] \, 
{\cal L}_{\mm  \mm} \mid ~~+  
\cr
&~~~~~~~ \fracm 12 \int d^2 \s \, d^{2} {\Bar{\zeta}}^{\pp} ~ {\Bar{\cal E}}^{-1} 
\left[\,  {\frac12} C^{ij} \, \nabla_{+ i} \nabla_{+ j} \, \right] {\cal L}_{\mm 
\mm} \mid  
~~~.  } \eqno(20) $$
Thus, we find that the density multiplet formulae provide a simple prescription 
for deriving locally (4,0) supersymmetrically invariant component actions from 
superspace actions. 

It is well known that the critical dimension of (4,0) strings is such that
classical conformal invariance does not survive quantization.  Even so, we
now have a lot of experiences to indicate that there are still interesting
phenomena occurring within such theories. The fact that there are four
different (4,0) scalar multiplets adds an extra twist...there are four candidates
from which to start.  These are the local versions of the actions in
(13)
$$ {\cal S}_{\rm {SM-I}} ~=~ \Big[ ~\int d^2 \s \, d^2 \z^{\pp} \, {\cal E}^{-1}~ 
\{ - \, i \frac 14 {\Bar {\cal B}} \, \nabla_= {\cal A} \, \} ~+~ {\rm h.} 
{\rm c.}  ~ \Big] ~~~~, 
\eqno(21) $$
$$ {\cal S}_{\rm {SM-II}} ~=~ \Big[ ~\int d^2 \s \, d^2 \z^{\pp} \, {\cal 
E}^{-1} ~  [~ - \, \frac 12 \Psi_{- i} \{ \, \nabla_{\mm} {\Bar \l}^{- i} \, 
+ \,  \frac 12 ( \S^{+ \, i} \phi ~+~ i \S^{+ \, j} \phi_j {}^i ) \, \}
~] ~+~ {\rm h.} 
{\rm c.} ~ \Big] ~~~~, 
\eqno(22) $$
$$ \eqalign{
{\cal S}_{\rm {SM-III}} ~=~ \Big[ ~\int d^2 \s \, d^2 \z^{\pp} \, {\cal E}^{-1}
~ [~ &- \, \frac 18 \S_- \{ \, \nabla_{\mm} {\Bar \pi}^-  ~+~  i \frac 18 C_{i 
\, j} \S^{+ \, i} {\Bar {\cal A}}^j \, \}   \cr
& ~+~ \frac 18 \U_- \{ \, \nabla_{\mm}  {\rho}^- ~-~ i 
\frac 12 \S^{+ \, i} {\cal A}_i \, \}
~]  ~+~ {\rm h.} {\rm c.} ~ \Big] ~~~~.  }
\eqno(23) $$

Here it is appropriate to make comments on these action formulae as well as
that of SM-IV. The actions above are found by beginning with the rigid results
and demanding the existence of their local extensions. In particular, the
chirality requirement of the integrands demands the appearance of the (4,0)
supergravity field strength supertensor. We thus find non-minimal coupling 
to the supergravity fields. These are explicitly seen for SM-II and SM-III 
theories. However, no non-minimal coupling is required for SM-I.  These 
results illustrate the ``unknown'' theorem that we have noted several times 
previously \cite{C}, \cite{Nish}, \cite{JMM}.   Namely, the result that the 
coupling to SM-I is minimal corresponds to the component level statement that 
the spin-0 fields in the SM-I multiplet are singlets under the (4,0) superholonomy 
group.  Note that the non-minimal coupling is such that only the supergravity-SM-I 
system possesses the full (4,0) superconformal invariance required of a 
string theory!

The reader will note that we have not presented a local extension for the
SM-IV theory. The reason for this is that at present there are still some
aspects of this theory that are being studied further.   We hope to report 
on this in the near future.

\noindent
{\bf{(V.) (4,0) Supergravity Coupled to Minimal (4,0) Scalar Multiplets:
\newline ${~~~\,~~}$ Component Results}}

Having derived the superspace form of the local versions of three of
our four multiplets, we wish to investigate the component results that
follow as consequences.  The distinctiveness of each multiplet will 
be crystal clear as a result.  All of our results below follow from 
the straightforward application of the density projectors developed 
in the previous section.

The SM-I multiplet has the following locally supersymmetrically
invariant action,
\begin{eqnarray}
{\cal S}_{\rm {SM-I}} & = & \int d^{2}\sigma \, e^{-1} ~ \Big[ ~~
                      \frac{1}{2} g^{m n} \{ (\pa_m  \cab)(\pa_n \ca) \, + \,
                                     (\pa_m  \cbb)(\pa_n \cb) \}
                                                   \nonumber \\
        &  &{~~~~~~~~~~~~~~~~} -i \{ {\sib i}\cdm{\si i} - (\cdm{\sib i}
             ){\si i} \}
                                                   \nonumber \\
        &  &{~~~~~~~~~~~~~~~~} + 2 (\cdp\cab)\{ C_{ij}{\sm i}{\si j} \}
             - 2 (\cdp\ca) \{ C^{ij}{\smb i}{\sib j} \}
                                                   \nonumber \\
        &  &{~~~~~~~~~~~~~~~~} + 2i (\cdp\cbb)\{ {\smb i}{\si i} \}
             + 2i (\cdp\cb) \{ {\sm i}{\sib i} \}
                                                   \nonumber \\
        &  &{~~~~~~~~~~~~~~~~} - 2 \{ {\sp i}{\sib i} \}\{ {\smb j}{\si j} \}
                               - 2 \{ {\sm i}{\sib i} \}\{ {\spb j}{\si j} \}
                                                   \nonumber \\
        &  &{~~~~~~~~~~~~~~~~} - 2 \{ C_{ij}{\sp i}{\si j} \}\{ C^{kl}{\smb k}
                                  {\sib l} \}
                                                    \nonumber \\
        &  &{~~~~~~~~~~~~~~~~} - 2 \{ C_{ij}{\sm i}{\si j} \}\{ C^{kl}{\spb k}
                              {\sib l} \}  ~~  \Big] ~~~.
                                                   \nonumber
\end{eqnarray}
$${~~}  \eqno(24)$$
\noindent
In the case of the SM-II theory the component result takes the form,
\begin{eqnarray}
{\cal S}_{\rm {SM-II}} & = & \int d^{2}\sigma \, e^{-1} ~\Big[ ~\frac{1}{2}(
                       \cdp\f)  (\cdm\f) + \frac{1}{4}(\cdp{\fai ij})
                       (\cdm{\fai ji}) {~~~~~~~~}
                                                         \nonumber \\
         &   &{~~~~~~~~~~~~~~~~} -   \frac{i}{2} \{ {\lmb i}\cdm{\lm i} 
                                 - (\cdm{\lmb i}){\lm i} \}
                                 + \frac{1}{4} {\fai ij}{\CF ji}\f       
                                                         \nonumber \\
         &   &{~~~~~~~~~~~~~~~~}  - \frac{1}{4} \{ \f^{2} - \frac{1}{2}
                                 {\fai ij}{\fai ji} \} \{ \frac{1}{2}\car 
                                 - i{\sp k}{\sigb k} - i{\spb k}{\sig k} \}
                                                          \nonumber \\
         &   &{~~~~~~~~~~~~~~~~} - \{ {\lm i}{\sm j} - {\lmb j}{\smb i} \}
                                 \cdp{\fai ji} - \frac{1}{2} \{ {\sig j}
                                {\lm i} - {\sigb i}{\lmb j} \}{\fai ji}
                                                          \nonumber \\
         &   &{~~~~~~~~~~~~~~~~} - i\{ {\lm i}{\sm i} + {\lmb i}{\smb i} 
                                \}\cdp\f + \frac{i}{2} \{ {\sig i}{\lm i} 
                                + {\sigb i}{\lmb i} \}\f
                                                          \nonumber \\
         &   &{~~~~~~~~~~~~~~~~} + \{ C^{ij}{\lm i}{\lm j} \}\{ C_{kl}{\sm 
                                   k}{\sp l} \}
                                                          \nonumber \\
         &   &{~~~~~~~~~~~~~~~~}+ \{ C_{ij}{\lmb i}{\lmb j} \}\{ C^{kl}
                                {\smb k}{\spb l} \}
                                                          \nonumber \\
         &   &{~~~~~~~~~~~~~~~~} - \{ {\lmb i}{\lm i} \}\{ {\sm j}{\spb j} 
                                  + {\sp j}{\smb j} \}
                           ~~ \Big] ~~~.       \nonumber
\end{eqnarray}
$${~~}  \eqno(25)$$
In the case of the SM-III theory the component result takes the form,
$$ \eqalign{
{\cal S}_{{\rm {SM-III}}}  &=~ \int d^{2}\sigma \, e^{-1}  ~ \Big[ \,
                \frac{1}{4} g^{m n} ({\cal D}_m {\aib i})({\cal D}_n {\ai i})    
                                                      \cr
    &   {~~~~~~~~~~~~~~~~~~~~} - \frac{i}{8} \{ \pmb\cdm\pm - (\cdm\pmb)\pm \}  
                            - \frac{i}{8} \{ {\overline \rho}^-\cdm{\rho}^- 
                            - (\cdm{\overline \rho}^-){\rho}^- \}
                                                      \cr
    &   {~~~~~~~~~~~~~~~~~~~~} - \frac{1}{2} \{ {\sm i}{\overline \rho}^- - C^{ij}
                            {\smb j}\pmb \}  \cdp{\ai i}   \cr
    &   {~~~~~~~~~~~~~~~~~~~~} + \frac{1}{2} \{ {\smb i}{\rho}^- - C_{ij}{\sm j}
                            \pm \}\cdp{\aib i}
                                                      \cr
    &   {~~~~~~~~~~~~~~~~~~~~} + \frac{1}{8} {\aib i}{\ai i} \{ \frac{1}{2}\car -
                            i{\sp j}{\sigb j} - i{\spb j}{\sig j} \}
                                                       \cr
    &   {~~~~~~~~~~~~~~~~~~~~} + \frac{1}{8} C_{ij}{\aib i}{\sig j}\pm
          - \frac{1}{8} C^{ij}{\ai i}{\sigb j}\pmb 
          + \frac{1}{8} {\sig i}{\ai i}{\overline \rho}^-
          - \frac{1}{8} {\sigb i}{\aib i}{\rho}^-
                                                      \cr
    &   {~~~~~~~~~~~~~~~~~~~~} - \frac{1}{4} \{ {\sm i}{\spb i} + {\sp i}{\smb i} \}
                           \{ \pmb\pm - {\overline \rho}^-{\rho}^- \}
                                                      \cr
    &   {~~~~~~~~~~~~~~~~~~~~} + \frac{1}{2} \{ C_{ij}{\sm i}{\sp j} \}\{ 
                            \pm{\overline \rho}^- \}  } $$
$$   {~} + \frac{1}{2} \{ C^{ij}{\smb i}{\spb j} \}\{ 
                            \pmb{\rho}^- \}
                                 ~~ \Big] ~~~.  {~~~~~~}   \eqno(26)$$

In order to simplify the subsequent discussion, let us set all purely fermionic
terms to zero to obtain
$$
{\cal S}_{\rm {SM-I}}  ~=  \int d^{2}\sigma \, e^{-1} ~ \Big[ ~
 \frac{1}{2} g^{m n} \{ \, (\pa_m \cab)(\pa_n \ca) \, + \, 
(\pa_m {{\Bar {\cal B}}})(\pa_n {{\cal B}}) \} ~~ \Big] ~~~,  
\eqno(27)$$
$$
\eqalign{
{\cal S}_{\rm {SM-II}}   &=  \int d^{2}\sigma \, e^{-1} ~\Big[ ~\frac{1}{2}(
                       \cdp\f)  (\cdm\f) + \frac{1}{4}(\cdp{\fai ij})
                       (\cdm{\fai ji}) {~~~~~~~~~~~~~~} \cr
         &{~~~~~~~~~~~~~~~~~~~~~}  + \frac{1}{4} {\fai ij}{\CF ji}\f 
                                  - \frac{1}{8} \, \{ \f^{2} - \frac{1}{2}
                                 {\fai ij}{\fai ji} \} \car 
                                   ~~ \Big] ~~~, }  \eqno(28) $$
$$
{\cal S}_{{\rm {SM-III}}}  = \int d^{2}\sigma \, e^{-1}  ~ \Big[ \,
 \frac{1}{4} g^{m n} ({\cal D}_m {\aib i})({\cal D}_n {\ai i}) 
\, +  \, \frac{1}{16} {\aib i}{\ai i} \car 
                          ~ \Big] ~~~.
  \eqno(29)$$
Note that SM-I possesses a completely conformal coupling of the world 
sheet zweibein to the spin-0 fields of the matter multiplet. For SM-II, 
the world sheet curvature ($\car$) as well as the SU(2) field strength
(${\CF ij})$ are both coupled to the spin-0 fields. This is in addition
to the implicit minimal SU(2) gauge field coupling inside the covariant
derivatives.  Finally for SM-III we see only the world sheet curvature
as an explicit coupling to the spin-0 fields as well the implicit
coupling to the SU(2) gauge fields via the covariant derivatives.

\noindent
{\bf {(VI.)  Discussion}}

One of the interesting consequences of our study of local (4,0) actions is
that it permits us to answer questions that were raised in the immediate past.
It was noticed that the phenomenon of a multiplicity of scalar multiplets 
also occurs in full (4,4) theory.  This led to our asking the natural 
questions \cite{GT}, ``Why are there so many $N$ = 4 superstrings?'' and ``How 
many $N$ = 4 superstrings exist?'' In fact, we found evidence of {\underline {
eight}} 2D (4,4) hypermultiplets \cite{C}.  On the basis of our study 
of (4,0) models, the answer to these questions are, ``Parity and four,
respectively'' In the notation of \cite{C} these $N$ = 4 superstrings 
are based on the $4s^+ $, $3s^+ s^- $ and $2s^+ 2s^- $ hypermultiplets\footnote{
We have noted previously that there is a two-fold degeneracy in the $2s^+ 2s^- $ 
case.}. The concept of distinct extended superstrings for a fixed value of 
$N$ may be new to some of our readers. So it may useful to review the first 
discovery of this phenomenon in a simpler context and use some concepts from 
superconformal field theory.

A number of years ago \cite{GLO} it was pointed out that within the context of 
$N$ = 2 superstrings, there {\underline {must}} exist a minimum of {\underline 
{three}} {\underline {distinct}} theories!  This was based on the fact that 
more than one type of $N$ = 2 scalar multiplet was known to exist. There is 
the standard 2$D$, $N$ = 2 chiral scalar multiplet as well as the 2$D$, $N$ = 
2 twisted chiral scalar multiplet. Either of these two scalar multiplets can
be used to write anomaly-free 2$D$, $N$ = 2 superstrings and there are
three possible ways to carry out such a construction.  We shall call these 
the $C^2$, $CT$ and $T^2$ $N$ = 2 superstrings.  The existence of both
the chiral scalar multiplet and the twisted chiral scalar multiplet are a
reflection of the fact that both (c,c) and (a,c) rings exist within
2$D$, $N$ = 2 superconformal field theory. The former correspond to chiral
superfields while the latter correspond to twisted chiral superfields.
Thus, in the construction of 2$D$, $N$ = 2 superstrings, there is one
version where the matter superfields possess mirror symmetry (the $CT$
$N$ = 2 superstring) if we neglect supergravity and two versions that are 
mirror asymmetric (the $C^2$ and $T^2$ $N$ = 2 superstrings).  If we neglect 
supergravity, these latter two theories are the ``mirror reflections'' of 
each other. A fundamental difference between a chiral and twisted
chiral multiplet is that the spin-0 states of the former have the
same parity while those of the latter have opposite parities.

Returning now to the $N$ = 4 case, we obtain the number four due to the
following implication of our work. For the (4,0) superstrings, we saw 
that in coupling the scalar multiplets to supergravity a very interesting
phenomenon occurred. Namely, whenever the scalar fields transformed 
non-trivially under the SU(2) of (4,0) supergravity, the locally
supersymmetric action demanded the presence of non-conformal couplings
in the world sheet action!  In other words, terms linear in the world sheet
curvature and quadratic in spin-0 fields are present unless the spin-0
fields were SU(2) singlets. This same phenomenon must occur in the full
(4,4) candidate superstring actions!  It is only in the case of the
$4s^+ $, $3s^+ s^- $ and $2s^+ 2s^- $ 2D (4,4) hypermultiplets that the
spin-0 fields are in the trivial representation of the SU(2) that is
gauged by (4,4) supergravity!  Thus a classification of the presently
known distinct 2$D$, $N$ = 4 superstrings consists of the $4s^+ $, $3s^+ 
s^- $, $2s^+ 2s^- {\rm A}$ and $2s^+ 2s^- {\rm B}$ superstrings. In
otherword, $N$ = 4 superstrings exist with either zero, one or two 
psuedo-scalar spin-0 fields replacing the usual scalar spin-0 fields. 
These are the direct generalizations of the analogous $N$ = 2 results 
(i.e. all of these are connected by different parity twists) and show 
that there is intrinsic non-uniqueness in $N$ = 4 superstring theory 
(exactly like $N$ = 2 theory) contrary to other suggestions \cite{BkVf}.
Stated another way, it is not $N$ = 4 superstrings that are unique
but instead it is the (4,0) superstring that presently seems unique.  
One final point is that the existence of both 2D, $N$ = 2 chiral and 
twisted chiral superfields are likely to be intimately tied to the 
existence of mirror symmetry.  Since we now know that suitable $N$ = 4 
parity twists exist, it is a natural question to wonder about $N$ = 4 
generalizations of mirror symmetry that might occur in some suitable 
systems.

Our observation regarding the $N$ = 4 superstring ``SU(2) singlet rule'' 
likely has one other unsettling implication. Some time ago \cite{PvaN}, 
a component level action was purportedly given for the 2D, $N$ = 4 
superstring.  Following that, we asserted  \cite{GTPvn} the equivalence 
of our superspace construction in \cite{GLO} to the prior work of
Pernici and van Nieuwenhuizen. It now appears that our assertion was
wrong.  The work of ref. \cite{PvaN} does not describe one of the 
twisted hypermultiplets, the work in ref. \cite{GLO} does.

In any event, the present work provides complete support for our 
interpretation that SM-I, SM-II and SM-III are distinct representations.
This is particularly obvious in the case of SM-I versus the other 
two multiplets considered here. If the claim that all the multiplets
in Table I are equivalent were true, then using field redefinitions
a conformal theory could be turned into a non-conformal theory!  We
don't believe that even the most misguided ``experts'' would make such 
a claim. We thus end with the following canticle, ``There are four
distinct minimal (4,0) scalar multiplets.''
$${~~~}$$

``{\it {Ye can lead a man up to the university but ye can't make him 
think.}}''  \newline ${~~~\,~~}$ -- Finley Peter Dunne

\newpage

\noindent
{\bf {Appendix: Results for Component Projection}}

In this appendix, we collect in one place the key results needed to
derive our component results from our superspace ones.

\begin{itemize}
\item Covariant Derivatives:
\begin{enumerate}
\item $\nabla_{\pp}| = \cdp + \sp{i}\pa_{+i} + \spb{i}{\Bar\pa}_{+}{}^{i}$
\item $\nabla_{=}| = \cdm + \sm{i}\pa_{+i} + \smb{i}{\Bar\pa}_{+}{}^{i}$ 
\item $\cdp = \ep + \op\cm + i{\cap ij}{\CY ji}$
\item $\cdm = \em + \om\cm + i{\cam ij}{\CY ji}$
\end{enumerate}
\item Spin Connections:
\begin{enumerate}
\item $\op = C_{\pp}{}_{\mm}{}^{\mm}$
\item $\om = - C_{\mm}{}_{\pp}{}^{\pp} + 2i \{ {\sp i}{\smb i}
             - {\sm i}{\spb i} \}$
\end{enumerate}
\item Field Strengths:
\begin{enumerate}
\item $-\frac{1}{2}{\sig i} = \cdp{\sm i} - \cdm{\sp i} + 2i \{
                             {\sp j}{\smb j} - {\sm j}{\spb j} \}{\sp i}$
\item $-\frac{1}{2}{\sigb i} = \cdp{\smb i} - \cdm{\spb i} + 2i \{
                             {\sp j}{\smb j} - {\sm j}{\spb j} \}{\spb i}$
\item $-\frac{1}{2}\car = \cdp\om - \cdm\op - i {\sp i}{\sigb i}
                          -i {\spb i}{\sig i} + 2i \{
                             {\sp i}{\smb i} - {\sm i}{\spb i} \}\op$
\item $-\frac{1}{2}{\CF ij} = \cdp{\cam ij} - \cdm{\cap ij} +
         \{ {\sp j}{\sigb i} - \frac{1}{2}{\del ij}{\sp k}{\sigb k} \} $
         \\ \hspace*{1.5cm}
         $ - \, \{ {\spb i}{\sig j}- \frac{1}{2}{\del ij}{\spb k}
         {\sig k} \} \, + \, 2i \{ {\sp k}{\smb k} - {\sm k}{\spb k} 
         \}{\cap ij}$
\end{enumerate}
\end{itemize}

\newpage

\end{document}